\documentclass[prd,twocolumn,tightenlines,showpacs,nofootinbib,superscriptaddress]{revtex4-1} 

\usepackage{amsfonts}
\usepackage{dcolumn}
\usepackage{bm}
\usepackage{amsmath}
\usepackage{nicefrac}
\usepackage{amsthm}
\usepackage{amssymb}
\usepackage{graphicx}
\usepackage{color}
\usepackage{url}
\usepackage{cancel}
\usepackage{rotating}

\usepackage[colorlinks=true,linkcolor=black,citecolor=blue,urlcolor=blue,bookmarksopen]{hyperref}

\newcommand{\appropto}{\mathrel{\vcenter{
  \offinterlineskip\halign{\hfil$##$\cr
    \propto\cr\noalign{\kern2pt}\sim\cr\noalign{\kern-2pt}}}}}
\renewcommand{\v}[1]{\boldsymbol{#1}}		

\begin{document}

\title{Improved limits on axion-like-particle-mediated P,T-violating interactions between electrons and nucleons from electric dipole moments of atoms and molecules}

\date{\today}
\author{Y.~V.~Stadnik} 
\affiliation{School of Physics, University of New South Wales, Sydney 2052, Australia}
\affiliation{Johannes Gutenberg University of Mainz, 55128 Mainz, Germany}
\author{V.~A.~Dzuba} 
\affiliation{School of Physics, University of New South Wales, Sydney 2052, Australia}
\author{V.~V.~Flambaum} 
\affiliation{School of Physics, University of New South Wales, Sydney 2052, Australia}
\affiliation{Johannes Gutenberg University of Mainz, 55128 Mainz, Germany}

\begin{abstract}
In the presence of P,T-violating interactions, the exchange of axion-like particles between electrons and nucleons in atoms and molecules induces electric dipole moments (EDMs) of atoms and molecules. 
We perform calculations of such axion-exchange-induced atomic EDMs using the relativistic Hartree-Fock-Dirac method including electron core polarisation (RPA) corrections. 
We present analytical estimates to explain the dependence of these induced atomic EDMs on the axion mass and atomic parameters. 
From the experimental bounds on the EDMs of atoms and molecules, 
including $^{133}$Cs, $^{205}$Tl, $^{129}$Xe, $^{199}$Hg, $^{171}$Yb$^{19}$F, $^{180}$Hf$^{19}$F$^+$ and $^{232}$Th$^{16}$O, 
we constrain the P,T-violating scalar-pseudoscalar nucleon-electron and electron-electron interactions mediated by a generic axion-like particle of arbitrary mass. 
Our limits improve on existing laboratory bounds from other experiments by many orders of magnitude for $m_a \gtrsim 10^{-2}~\textrm{eV}$. 
We also place constraints on CP violation in certain types of relaxion models. 
\end{abstract}


\maketitle

\textbf{Introduction.} --- 
The Standard Model (SM) of particle physics has to date provided a very successful framework for describing and explaining most of the observed physical processes and phenomena in nature. 
However, despite its success, the SM does not explain several important observed phenomena, including dark matter and the observed matter-antimatter asymmetry in our Universe. 
This suggests the existence of new particles, which may interact feebly with the known particles of the SM, as well as additional sources of \emph{CP} violation beyond the SM. 

The axion, an odd-parity spin-0 particle that was originally proposed to resolve the strong \emph{CP} problem of Quantum Chromodynamics (QCD) \cite{PQ1977A,Weinberg1978,Wilczek1978,Kim1979,Zakharov1980,Zhitnitsky1980,Srednicki1981} and later realised to also be an excellent candidate for dark matter \cite{Preskill1983cosmo,Sikivie1983cosmo,Dine1983cosmo}, is a prominent example of such a particle \cite{Footnote1}. 

One may write the couplings of the QCD axion $a$ with the SM fermions $\psi$ in the following form:
\begin{equation}
\label{general_formula_SP}
\mathcal{L}_\textrm{int} = a \sum_\psi \bar{\psi} \left(g_\psi^s + ig_\psi^p \gamma_5 \right) \psi \, .
\end{equation}
In the absence of \emph{CP} violation in the QCD sector (i.e., when the QCD vacuum angle $\theta$ in the Lagrangian $\mathcal{L}_{\theta} = \theta g^2 G\tilde{G}/32\pi^2$ attains its minimum at $\theta_{\textrm{eff}} = 0$), the couplings of the axion with fermions are \emph{CP} conserving:~$g_\psi^s = 0$. 
However, when $\theta_{\textrm{eff}} \ne 0$, the axion acquires non-zero \emph{CP}-violating couplings with the light quarks:~$g_u^s = g_d^s = g_s^s = (\theta_{\textrm{eff}} m_u m_d) / [(m_u+m_d) f_a]$, where $f_a$ is the axion decay constant \cite{Wilczek1984}, and the subscripts $u$, $d$ and $s$ refer to the up, down and strange quark flavours, respectively. 
In this case, electric dipole moment (EDM) experiments with ultracold neutrons \cite{Baker2006,Pendlebury2015} and atomic mercury \cite{Heckel2016Hg}, which constrain the effective QCD vacuum angle to be $|\theta_{\textrm{eff}}| \lesssim 10^{-10}$, place the following bounds on the combination of parameters $g_q^s g_\psi^p$ (here $\psi$ denotes either a light quark or the electron, with $g^p_\psi = m_\psi/f_a$): 
\begin{equation}
\label{QCD_axion_bounds}
|g_q^s g_\psi^p| \sim \frac{m_q |\theta_{\textrm{eff}}|}{f_a} \frac{m_\psi}{f_a} ~=>~ \frac{|g_q^s g_\psi^p|}{m_a^2} \lesssim \frac{10^{-10} m_q m_\psi}{\Lambda_{\textrm{QCD}}^4} \, , 
\end{equation}
where we have made use of the relation $m_a f_a \sim \Lambda_{\textrm{QCD}}^2$ for the QCD axion, with $\Lambda_{\textrm{QCD}} \sim 250~\textrm{MeV}$ being the QCD scale. 

Apart from the QCD axion, one may also consider generic axion-like particles, for which the contributions to $g_\psi^s$ are unrelated to the QCD sector, and so to which the bounds in Eq.~(\ref{QCD_axion_bounds}) do not apply. 
Indeed, the majority of searches for the \emph{CP}-violating couplings in Eq.~(\ref{general_formula_SP}) via the \emph{P},\emph{T}-violating interactions which they mediate make no specific assumption about the underlying source of \emph{CP} violation \cite{Wilczek1984,Bouchiat1975,Wineland1991,Venema1992,Ritter1993,Youdin1996,Pospelov1998,Ni1999,Heckel2006,Baessler2007,Hammond2007,Serebrov2009,Ignatovich2009,Serebrov2010,Petukhov2010,Hoedl2011,Raffelt2012,Tullney2013,Chu2013,Bulatowicz2013,Musolf2014,Geraci2014,Stadnik2015NMBE,Afach2015,Heckel2015,Ruoso2017,Rong2017}.

\begin{table*}[t]
\centering
\caption{
Summary of relativistic Hartree-Fock-Dirac calculations 
of the atomic EDMs induced by interaction (\ref{relativistic_potential_FULL}) for various axion masses. 
The presented values for the atomic EDMs are in terms of the parameter $C_{\textrm{SP}}^{(12)} = -\sqrt{2} g_1^p g_2^s / G_F  m_a^2$ and in the units $e \cdot \textrm{cm}$. 
For the electron-nucleon interaction, the values are normalised to a single nucleon, while for the electron-electron interaction, the values include the effects of all atomic electrons. 
For molecular YbF in the $^2 \Sigma_{1/2}$ state, we calculate $D = \left< s_{1/2} \left| d_z \right|  s_{1/2} \right> \equiv D (s_{1/2})$ for the Yb$^+$ ion. 
For molecular HfF$^+$ and ThO in the $^3 \Delta_1$ excited metastable state, we calculate $D = - D(s_{1/2}) + \frac{3}{5} D(d_{5/2})$ for the Hf$^{3+}$ and Th$^+$ ions, respectively. 
}
\label{tab:table1}
\resizebox{\linewidth}{!}{
\begin{tabular}{ |c|c|c|c|c|c|c|c|c|c|c| }%
\hline
 & \multicolumn{2}{c}{Cs} & \multicolumn{2}{|c|}{Tl} & \multicolumn{2}{c}{Yb$^+$} & \multicolumn{2}{|c|}{Hf$^{3+}$} & \multicolumn{2}{c|}{Th$^{+}$}  \\ \hline
$m_a$~(eV) & $d_{\textrm{a}}/C_\textrm{SP}^{(eN)}$ & $d_{\textrm{a}}/C_\textrm{SP}^{(ee)}$ & $d_{\textrm{a}}/C_\textrm{SP}^{(eN)}$ & $d_{\textrm{a}}/C_\textrm{SP}^{(ee)}$ & $D/C_\textrm{SP}^{(eN)}$ & $D/C_\textrm{SP}^{(ee)}$ & $D/C_\textrm{SP}^{(eN)}$ & $D/C_\textrm{SP}^{(ee)}$ & $D/C_\textrm{SP}^{(eN)}$ & $D/C_\textrm{SP}^{(ee)}$  \\ \hline 
$\infty$ & $+7.7 \times 10^{-19}$ & $+4.4 \times 10^{-20}$ & $-7.1 \times 10^{-18}$ & $-2.0 \times 10^{-19}$ & $+2.0 \times 10^{-18}$ & $+7.9 \times 10^{-20}$ & $-2.3 \times 10^{-18}$ & $-8.8 \times 10^{-20}$ & $-5.8 \times 10^{-17}$ & $-1.4 \times 10^{-18}$ \\ \hline
$10^8$ & $+7.4 \times 10^{-19}$ & $+4.4 \times 10^{-20}$ & $-6.7 \times 10^{-18}$ & $-2.0 \times 10^{-19}$ & $+1.8 \times 10^{-18}$ & $+8.0 \times 10^{-20}$ & $-2.2 \times 10^{-18}$ & $-8.9 \times 10^{-20}$ & $-5.4 \times 10^{-17}$ & $-1.5 \times 10^{-18}$ \\ \hline
$10^7$ & $+5.3 \times 10^{-19}$ & $+4.4 \times 10^{-20}$ & $-3.5 \times 10^{-18}$ & $-2.0 \times 10^{-19}$ & $+1.1 \times 10^{-18}$ & $+7.9 \times 10^{-20}$ & $-1.3 \times 10^{-18}$ & $-8.8 \times 10^{-20}$ & $-2.5 \times 10^{-17}$ & $-1.4 \times 10^{-18}$ \\ \hline
$10^6$ & $+1.9 \times 10^{-19}$ & $+2.9 \times 10^{-20}$ & $-5.9 \times 10^{-19}$ & $-5.1 \times 10^{-20}$ & $+2.6 \times 10^{-19}$ & $+3.6 \times 10^{-20}$ & $-2.9 \times 10^{-19}$ & $-3.8 \times 10^{-20}$ & $-3.1 \times 10^{-18}$ & $-2.8 \times 10^{-19}$ \\ \hline
$10^5$ & $+4.1 \times 10^{-21}$ & $-7.4 \times 10^{-21}$ & $-4.3 \times 10^{-21}$ & $+2.3 \times 10^{-20}$ & $+3.2 \times 10^{-21}$ & $-1.3 \times 10^{-20}$ & $-3.4 \times 10^{-21}$ & $+1.3 \times 10^{-20}$ & $-2.3 \times 10^{-20}$ & $+1.3 \times 10^{-19}$ \\ \hline
$10^4$ & $-6.5 \times 10^{-24}$ & $-9.1 \times 10^{-22}$ & $+1.1 \times 10^{-23}$ & $+1.5 \times 10^{-21}$ & $-6.9 \times 10^{-24}$ & $-9.5 \times 10^{-22}$ & $+7.7 \times 10^{-24}$ & $+1.0 \times 10^{-21}$ & $+6.4 \times 10^{-23}$ & $+1.0 \times 10^{-20}$ \\ \hline
$10^3$ & $-6.3 \times 10^{-25}$ & $-3.4 \times 10^{-23}$ & $+6.3 \times 10^{-25}$ & $+4.7 \times 10^{-23}$ & $-5.7 \times 10^{-25}$ & $-4.0 \times 10^{-23}$ & $+4.6 \times 10^{-25}$ & $+3.2 \times 10^{-23}$ & $+3.2 \times 10^{-24}$ & $+2.8 \times 10^{-22}$ \\ \hline
$10^2$ & $-8.3 \times 10^{-27}$ & $-4.4 \times 10^{-25}$ & $+6.6 \times 10^{-27}$ & $+4.9 \times 10^{-25}$ & $-6.7 \times 10^{-27}$ & $-4.7 \times 10^{-25}$ & $+5.1 \times 10^{-27}$ & $+3.5 \times 10^{-25}$ & $+3.9 \times 10^{-26}$ & $+3.4 \times 10^{-24}$ \\ \hline
$10$ & $-8.3 \times 10^{-29}$ & $-4.4 \times 10^{-27}$ & $+6.6 \times 10^{-29}$ & $+4.9 \times 10^{-27}$ & $-6.8 \times 10^{-29}$ & $-4.8 \times 10^{-27}$ & $+5.1 \times 10^{-29}$ & $+3.5 \times 10^{-27}$ & $+3.9 \times 10^{-28}$ & $+3.4 \times 10^{-26}$ \\ \hline

\end{tabular}
}
\end{table*}

In the present work, we investigate the manifestation of the exchange of generic axion-like particles of arbitrary mass between electrons and nucleons in atoms and molecules, in the presence of the couplings in Eq.~(\ref{general_formula_SP}). 
The \emph{P},\emph{T}-violating potential due to the exchange of an axion of mass $m_a$ between two fermions reads: 
\begin{equation}
\label{relativistic_potential_FULL}
V_{12}(r) = +i \frac{g_1^p g_2^s}{4 \pi} \frac{e^{-m_a r}}{r} \gamma^0 \gamma_5 \, ,
\end{equation}
where $r$ is the distance between the two fermions, and the $\gamma$-matrices correspond to fermion 1. 
We restrict our attention to the case when fermion 1 is the electron, but fermion 2 can be either the electron or nucleons. 
We also introduce the shorthand notation $g_N^s \equiv (N g_n^s + Z g_p^s) / A$, where $N$ is the neutron number, $Z$ is the proton number, and $A = Z + N$ is the nucleon number. 

The \emph{P},\emph{T}-violating potential in Eq.~(\ref{relativistic_potential_FULL}) induces EDMs in atoms and molecules by mixing atomic states of opposite parity. 
We perform calculations of such axion-exchange-induced atomic EDMs using the relativistic Hartree-Fock-Dirac method including electron core polarisation (RPA) corrections. 
We summarise our results in Tables~\ref{tab:table1} and \ref{tab:table2}. 
Detailed analytical calculations explaining the dependence of these induced atomic EDMs on the axion mass and atomic parameters are presented in the Supplemental Material.

\textbf{Calculations.} --- 

\emph{Paramagnetic atoms.} --- 
We perform calculations of axion-exchange-induced EDMs of paramagnetic atoms using the relativistic Hartree-Fock-Dirac method including electron core polarisation (RPA) corrections. 
For the atomic EDM of Tl, electron correlation corrections are known to play an important role (see, e.g., Ref.~\cite{Dzuba2009Cs+Tl}). 
Therefore, for Tl, we employ the CI+MBPT method described in \cite{Dzuba2009Cs+Tl} to perform the EDM calculations in the present work. 
Correlations between the core electrons and three valence electrons in Tl (ground state $6s^26p_{1/2}$) have been taken into account using the many-body perturbation theory (MBPT)  method including the screening of the valence electron interactions by the core electrons. 
The Hamiltonian matrix for the three valence electrons has been diagonalised using the configuration interaction (CI) approach. 

\emph{Paramagnetic molecules.} --- 
In molecular species, the heavy atom is in the internal electric field of a molecule, $\v{E}_{\textrm{int}}$, and so the corresponding energy shift may be estimated by $\Delta \varepsilon \approx - \v{D} \cdot \v{E}_\textrm{int}$, where $\v{D}$ is the induced EDM of the heavy atomic species. 
The molecular electric field cancels out in the ratio:
\begin{equation}
\label{molecule_assumption}
\frac{\Delta \varepsilon|_{m_a}}{\Delta \varepsilon|_{m_a \to \infty}} \approx \frac{D|_{m_a}}{D|_{m_a \to \infty}} \, ,
\end{equation}
where the subscripts refer to the axion masses at which the relevant quantities are evaluated. 
Expression (\ref{molecule_assumption}) allows us to determine the energy shift for a finite axion mass in molecules, by using calculated values for the induced EDM of the heavy atomic species in Table~\ref{tab:table1}, as well as existing values of the energy shift for an infinite axion mass in molecules \cite{Kozlov1997YbF,Parpia1998YbF,Mosyagin1998YbF,Quiney1998YbF,Skripnikov2017HfF+,Fleig2017HfF+,Petrov2007HfF+,Fleig2013HfF+,Meyer2008ThO,Skripnikov2013ThO,Fleig2014ThO}. 
This allows us to interpret molecular experiments.

For molecular YbF, which is in the $^2 \Sigma_{1/2}$ state, we calculate $D = \left< s_{1/2} \left| d_z \right|  s_{1/2} \right> \equiv D (s_{1/2})$ for the Yb$^+$ ion, where the atomic EDM is calculated for the maximal projection of the electron angular momentum, $j_z$. 
For molecular HfF$^+$ and ThO, the EDM is measured in the $^3 \Delta_1$ excited metastable state that corresponds to one $s$ and one $d$ electron in the state $\left| L_z=+2, S_z=-1, J_z = +1 \right>$. 
Expanding this state in terms of $s_{1/2}$, $d_{3/2}$ and $d_{5/2}$ atomic orbitals, we obtain $D = - D(s_{1/2}) + \frac{3}{5} D(d_{5/2})$. 
The $d_{3/2}$ atomic orbital does not contribute to the atomic EDM in this case, since the dipole operator cannot mix it with a $p_{3/2}$ atomic orbital (which has the opposite value of the electron spin projection, $s_z$). 

For a high-mass axion (where the effect arises mainly from the small distances $r \ll a_\textrm{B}/Z^{1/3}$), the dominant contribution to the atomic EDM comes from the mixing of the $s_{1/2}$ state with $p_{1/2}$ states, while for a low-mass axion (where the effect arises mainly from the intermediate distances $r \sim a_\textrm{B}/Z^{1/3}$), there is also a non-negligible contribution from the mixing of the $d_{5/2}$ state with $f_{5/2}$ states.

\begin{sidewaystable}
\centering
\caption{
Summary of derived limits on the combinations of parameters $g_N^s g_e^p X_r/m_a^2$ and $g_e^s g_e^p /m_a^2$ for $m_a \gg 300~\textrm{keV}$, and $g_N^s g_e^p$ and $g_e^s g_e^p$ for $m_a \ll 1~\textrm{keV}$, from the consideration of tree-level axion-mediated \emph{P},\emph{T}-violating interactions between electrons and nucleons in atoms and molecules, and on the combination of parameters $g_e^s g_e^p \ln(m_a/m_e) /m_a^2$ for $m_a \gg m_e$, from the consideration of the loop-induced electron EDM. 
The parameter $X_r$ is defined by $X_r \approx 1$ when $m_a R_{\textrm{nucl}} \gg 1$, and $X_r \approx (m_a R_{\textrm{nucl}})^{2-2\gamma}$ when $m_a R_{\textrm{nucl}} \ll 1$, where $R_{\textrm{nucl}}$ is the radius of the atomic nucleus and $\gamma = \sqrt{(j+1/2)^2 -(Z \alpha)^2}$, see the Supplemental Material for more details. 
We have also summarised the numerical calculations (see also Table~\ref{tab:table1}) and experimental EDM bounds used in deriving these limits. 
The \emph{P},\emph{T}-odd parameters $W_c$ and $W_d$ are the normalised expectation values of the contact nucleon-electron scalar-pseudoscalar interaction operator $H_{\textrm{SP}} = - i(G_F C_{\textrm{SP}} / \sqrt{2}) \gamma^0 \gamma_5 \delta^{(3)}(\v{r})$, and of the electron EDM interaction operator $H_{\textrm{e}} = -d_e \gamma^0 \v{\Sigma} \cdot \v{E}$, respectively:~$W_c = \left< \Psi \left| H_{\textrm{SP}} \right| \Psi \right>/  (C_{\textrm{SP}} \Omega)$, $W_d = \left< \Psi \left| H_{\textrm{e}} \right| \Psi \right>/ (d_{\textrm{e}} \Omega)$. 
The best limits are highlighted in bold. 
}
\label{tab:table2}
\resizebox{\linewidth}{!}{
\begin{tabular}{ c|c|c|c|c|c|c|c|c }%
Atom & $d_{\textrm{a}}/C_\textrm{SP}~(e \cdot \textrm{cm})$  & $d_{\textrm{a}}/d_{\textrm{e}}$ & $|d_{\textrm{a}}|$ exp.~limit $(e \cdot \textrm{cm})$ & $|g_N^s g_e^p|X_r/m_a^2$ limit $(\textrm{GeV}^{-2})$ & $|g_N^s g_e^p|~\textrm{limit}$ & $|g_e^s g_e^p|/m_a^2$ limit $(\textrm{GeV}^{-2})$ & $|g_e^s g_e^p| \ln(m_a/m_e) /m_a^2$ limit $(\textrm{GeV}^{-2})$ & $|g_e^s g_e^p|~\textrm{limit}$  \\ \hline 

$^{133}$Cs & $+7.6 \times 10^{-19}$ \cite{Dzuba2009Cs+Tl} & $+124$ \cite{Dzuba2009Cs+Tl} & $1.3 \times 10^{-23}$ \cite{Hunter1989Cs} & $1.5 \times 10^{-10}$ & $1.3 \times 10^{-16}$ & $3.4 \times 10^{-7}$ & $4.2 \times 10^{-7}$ & $3.3 \times 10^{-16}$  \\ \hline
$^{205}$Tl & $-7.0 \times 10^{-18}$ \cite{Dzuba2009Cs+Tl} & $-582$ \cite{Dzuba2009Cs+Tl,Kelly1992Tl,Kozlov2012Tl} & $9.4 \times 10^{-25}$ \cite{DeMille2002Tl} & $1.1 \times 10^{-12}$ & $1.2 \times 10^{-17}$ & $8.1 \times 10^{-9}$ & $6.3 \times 10^{-9}$ & $3.2 \times 10^{-17}$  \\ \hline
$^{129}$Xe & $-5.0 \times 10^{-23}$ \cite{Ginges2004Review} & $-8 \times 10^{-4}$ \cite{Ginges2004Review} & $6.6 \times 10^{-27}$ \cite{Chupp2001Xe} & $1.1 \times 10^{-9}$ & --- & $1.4 \times 10^{-6}$  & $3.2 \times 10^{-5}$ & ---  \\ \hline
$^{199}$Hg & $-5.9 \times 10^{-22}$ \cite{Ginges2004Review} & $-0.014$ \cite{Ginges2004Review} & $7.4 \times 10^{-30}$ \cite{Heckel2016Hg} & $1.0 \times 10^{-13}$ & --- & $5.5 \times 10^{-10}$  & $2.1 \times 10^{-9}$ & ---  \\ 
\multicolumn{1}{c}{} & \multicolumn{1}{c}{} & \multicolumn{1}{c}{} & \multicolumn{1}{c}{} & \multicolumn{1}{c}{} & \multicolumn{1}{c}{} & \multicolumn{1}{c}{} & \multicolumn{1}{c}{} & \multicolumn{1}{c}{} \\
\end{tabular}
}

\resizebox{\linewidth}{!}{
\begin{tabular}{ c|c|c|c|c|c|c|c|c }%
Molecule & $W_c/W_d~(e \cdot \textrm{cm})$ & $|\mathcal{E}_{\textrm{eff}}|~(\textrm{GV/cm})$ & $|d_{\textrm{e}}|$ exp.~limit $(e \cdot \textrm{cm})$ & $|g_N^s g_e^p|X_r/m_a^2$ limit $(\textrm{GeV}^{-2})$ & $|g_N^s g_e^p|~\textrm{limit}$ & $|g_e^s g_e^p|/m_a^2$ limit $(\textrm{GeV}^{-2})$ & $|g_e^s g_e^p| \ln(m_a/m_e) /m_a^2$ limit $(\textrm{GeV}^{-2})$ & $|g_e^s g_e^p|~\textrm{limit}$  \\ \hline
$^{171}$Yb$^{19}$F & $8.3 \times 10^{-21}$ \cite{Kozlov1997YbF} & $14.5$ \cite{Kozlov1997YbF,Parpia1998YbF,Mosyagin1998YbF,Quiney1998YbF,Hinds2011YbF} & $1.05 \times 10^{-27}$ \cite{Hinds2011YbF} & $1.0 \times 10^{-12}$ & $3.1 \times 10^{-18}$ & $4.5 \times 10^{-9}$ & $4.1 \times 10^{-9}$ & $7.5 \times 10^{-18}$  \\ \hline
$^{180}$Hf$^{19}$F$^+$ & $9.3 \times 10^{-21}$ \cite{Skripnikov2017HfF+,Fleig2017HfF+} & $23$ \cite{Petrov2007HfF+,Fleig2013HfF+} & $1.3 \times 10^{-28}$ \cite{Cornell2017HfFedm} & $1.2 \times 10^{-13}$ & $5.3 \times 10^{-19}$ & $5.5 \times 10^{-10}$ & $5.1 \times 10^{-10}$ & $1.4 \times 10^{-18}$  \\ \hline
$^{232}$Th$^{16}$O & $1.5 \times 10^{-20}$ \cite{Skripnikov2013ThO} & $84$ \cite{Meyer2008ThO,Skripnikov2013ThO,Fleig2014ThO} & $8.7 \times 10^{-29}$ \cite{DeMille2014ThO} & $4.8 \times 10^{-14}$ & $7.2 \times 10^{-19}$ & $4.5 \times 10^{-10}$ & $3.4 \times 10^{-10}$ & $1.9 \times 10^{-18}$  \\ \hline
$^{232}$Th$^{16}$O & $1.5 \times 10^{-20}$ \cite{Skripnikov2016ThO,Fleig2016ThO} & $78$ \cite{Skripnikov2016ThO,Fleig2016ThO} & $1.1 \times 10^{-29}$ \cite{ACME2018ThO} & $\mathbf{6.0 \times 10^{-15}}$ & $\mathbf{9.0 \times 10^{-20}}$ & $\mathbf{5.6 \times 10^{-11}}$ & $\mathbf{4.3 \times 10^{-11}}$ & $\mathbf{2.4 \times 10^{-19}}$  \\ 
\end{tabular}
}
\end{sidewaystable}

\emph{Diamagnetic atoms.} --- 
In diamagnetic atoms with zero electron angular momentum, an electron-spin-dependent \emph{P},\emph{T}-violating interaction induces an atomic EDM only in combination with the hyperfine interaction \cite{Flambaum1985EDM}. 
Calculations in Ref.~\cite{Flambaum1985EDM} have been performed for the contact limit of interaction (\ref{relativistic_potential_FULL}), and also for the interaction of an electron EDM with atomic electric and magnetic fields. 
Relativistic many-body calculations of the electron EDM effects including RPA corrections have been performed in Ref.~\cite{Martensson1987EDM}. 
There is an approximate analytical relation between the matrix elements of the contact limit of interaction (\ref{relativistic_potential_FULL}) and the interaction of an electron EDM with the atomic electric field \cite{Harabati2011EDM-relations}. 
Therefore, we may also use the calculations of the electron EDM effects to predict the effect of the contact limit of Eq.~(\ref{relativistic_potential_FULL}). 
In the present work, we use the calculated values for $d_{\textrm{a}}/C_\textrm{SP}$ (defined via the operator $H_{\textrm{SP}} = - i(G_F C_{\textrm{SP}} / \sqrt{2}) \gamma^0 \gamma_5 \delta^{(3)}(\v{r})$) from Refs.~\cite{Flambaum1985EDM,Martensson1987EDM}, which have been presented in the review \cite{Ginges2004Review}, together with the analytical formulae ({\color{blue}9}) and ({\color{blue}12}) in the Supplemental Material, in order to extract the limits presented in Table~\ref{tab:table2}.

\begin{figure}[h!]
\begin{center}
\includegraphics[width=8.5cm]{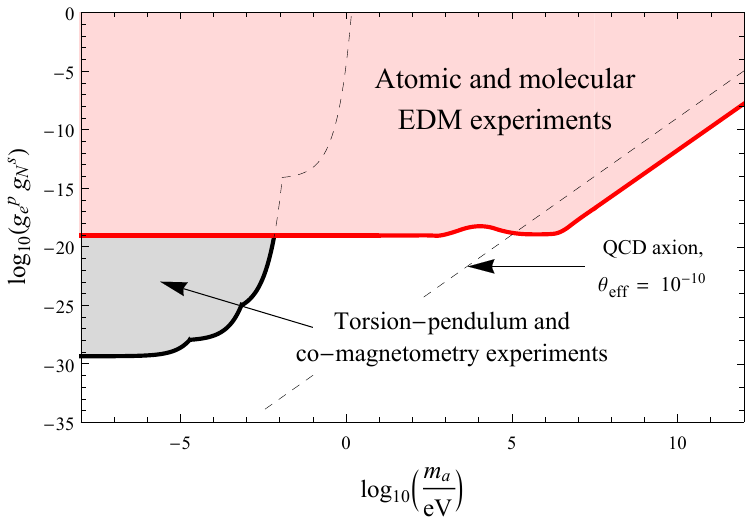} 
\includegraphics[width=8.5cm]{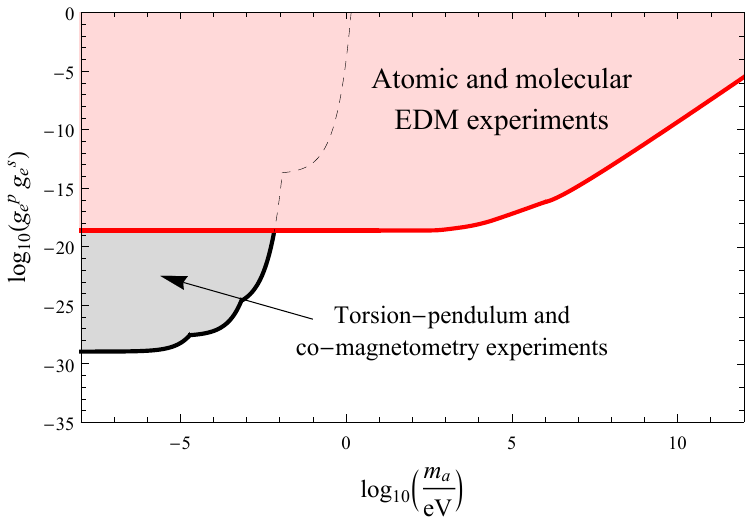} 
\caption{(Colour online)  Limits on the \emph{P},\emph{T}-violating scalar-pseudoscalar nucleon-electron (top figure) and electron-electron (bottom figure) interactions mediated by a generic axion-like particle of mass $m_a$, as defined in Eq.~(\ref{general_formula_SP}). 
The regions in red correspond to regions of parameters excluded by the present work from consideration of atomic and molecular electric dipole moment experiments. 
The regions in grey correspond to existing constraints from torsion-pendulum and co-magnetometry experiments \cite{Youdin1996,Ni1999,Hammond2007,Hoedl2011,Heckel2015,Ruoso2017,Rong2017}. 
The region above the dashed black line in the top figure corresponds to the limits for the QCD axion with $\left| \theta_\textrm{eff} \right| \lesssim 10^{-10}$. 
In extrapolating the limits on $g_e^s g_e^p$ from torsion-pendulum and co-magnetometry experiments using the published limits on $g_N^s g_e^p$ in \cite{Youdin1996,Ni1999,Hammond2007,Hoedl2011,Heckel2015,Ruoso2017,Rong2017}, we have assumed that $\bar{A} \approx 2.5\bar{Z}$ for the mean nuclear contents of the unpolarised test bodies. 
} 
\label{fig:Results_summary}
\end{center}
\end{figure}

\textbf{Results and Discussion.} --- 
Our results are summarised in Tables~\ref{tab:table1} and \ref{tab:table2}, and are shown in Fig.~\ref{fig:Results_summary}. 
We find that the best limits come from the newer ThO experimental data in Ref.~\cite{ACME2018ThO}, with quite strong limits on low-mass axions also coming from HfF$^+$. 
The reason why a relatively light system such as HfF$^+$ can give strong constraints for low-mass axions (and not necessarily for high-mass axions) can be traced to the dependence of the induced atomic EDM on the atomic parameters. 
When a high-mass axion is exchanged, the induced atomic EDM has a strong $Z$-dependence (scaling as $d_a \propto A Z^2 K_\textrm{rel}$ for the electron-nucleon interaction and $d_a \propto Z^2$ for the electron-electron interaction, where $K_\textrm{rel}$ is a relativistic factor), whereas when a low-mass axion is exchanged, the induced atomic EDM has a milder $Z$-dependence (scaling only as $d_a \propto A$ for the electron-nucleon interaction and $d_a \propto Z$ for the electron-electron interaction), see the Supplemental Material for more details. 

We also note that the atomic EDMs induced by the exchange of high-mass and low-mass axions differ in sign (see Table~\ref{tab:table1}). 
This can be traced to the fact that the effects arise from different distances in these two limiting cases. 
When a high-mass axion is exchanged, the dominant contribution comes from the small distances $r \ll a_\textrm{B}/Z^{1/3}$, whereas when a low-mass axion is exchanged, the dominant contribution comes from the intermediate distances $r \sim a_\textrm{B}/Z^{1/3}$, where the wavefunctions oscillate, see the Supplemental Material for more details.

\begin{figure}[h!]
\begin{center}
\includegraphics[width=3.5cm]{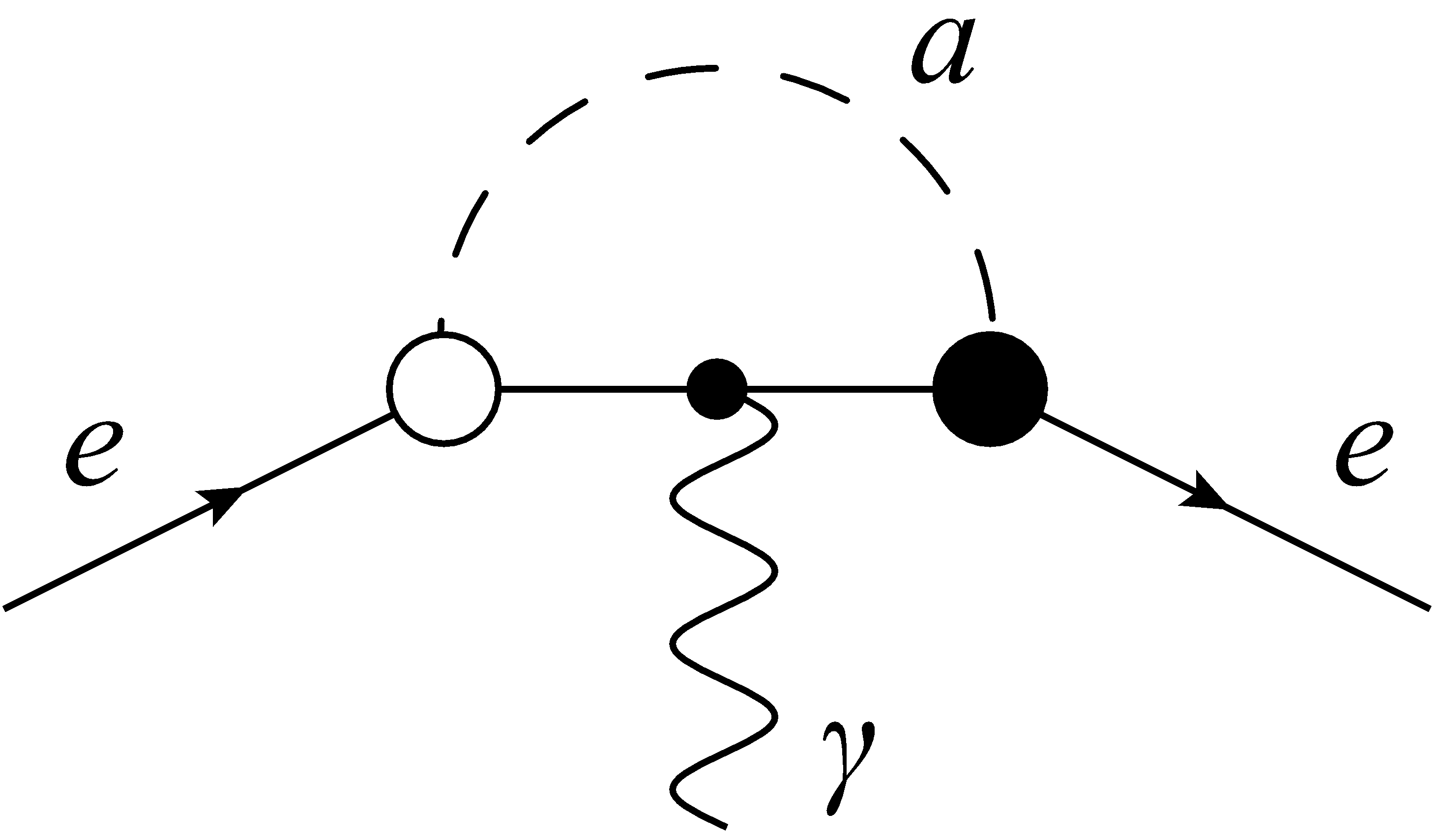} 
\includegraphics[width=3.5cm]{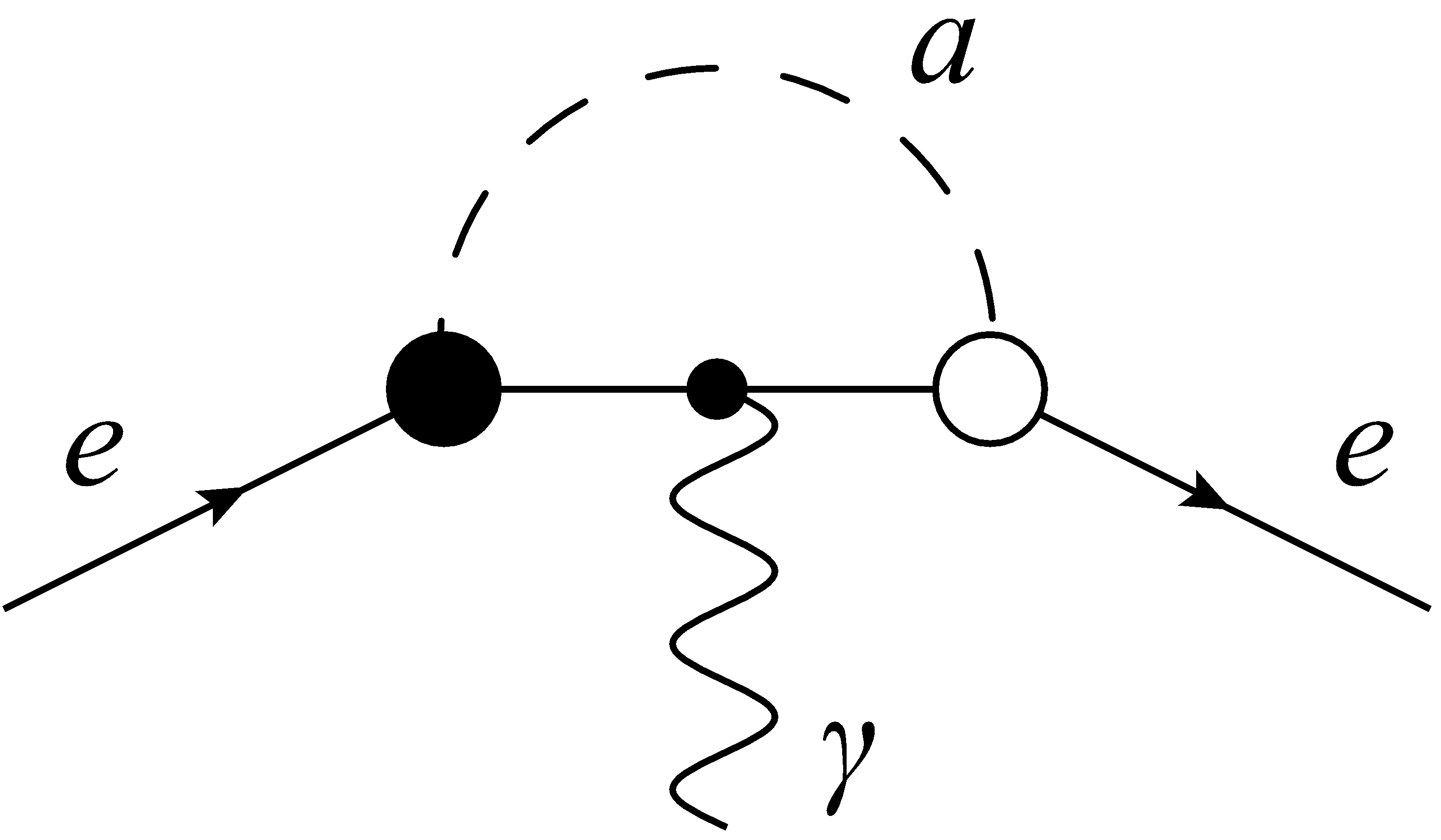} 
\caption{ 
1-loop-induced contribution to an electron electric dipole moment. 
The large black circle denotes a pseudoscalar interaction vertex, while the white circle denotes a scalar interaction vertex, as defined in Eq.~(\ref{general_formula_SP}). 
} 
\label{fig:loop_diagrams}
\end{center}
\end{figure}

\emph{Loop-induced electron EDM.} --- 
The interactions in Eq.~(\ref{general_formula_SP}) also induce an electron EDM via the 1-loop process in Fig.~\ref{fig:loop_diagrams}: 
\begin{align}
\label{electron_EDM_1loop}
&d_e \approx -\frac{g_e^s g_e^p e m_e}{4 \pi^2 m_a^2} \ln\left(m_a/m_e\right) &\textrm{for $m_a \gg m_e$,} \\
&d_e \approx -\frac{g_e^s g_e^p e}{8 \pi^2 m_e} &\textrm{for $m_a \ll m_e$,}
\end{align}
where $-e$ is the electric charge of the electron. 
Eq.~(\ref{electron_EDM_1loop}) was presented in \cite{Bouchiat1975}. 
We see (referring to the tabulated data in Tables~\ref{tab:table1} and \ref{tab:table2}) that the 1-loop-induced electron EDM contribution (proportional to $g_e^s g_e^p$) to the atomic and molecular EDMs is smaller than the corresponding direct tree-level contribution for small axion masses, but can be larger for large axion masses. 
The reason for the latter is the strong $Z$-dependence of the electron EDM contribution in species with unpaired atomic electrons ($d_\textrm{a} \propto Z^3 K_e d_e$, where $K_e$ is a relativistic factor \cite{Flambaum1976}), compared with that of the direct tree-level contribution ($d_a \propto Z^2 g_e^s g_e^p$), as well as an additional numerical suppression factor for the direct tree-level contribution (see Eq.~({\color{blue}11}) in the Supplemental Material).

\textbf{Conclusions.} --- 
To summarise, we have derived limits on the \emph{P},\emph{T}-violating scalar-pseudoscalar nucleon-electron and electron-electron interactions mediated by a generic axion-like particle of arbitrary mass from EDM experiments with atoms and molecules (see Table~\ref{tab:table2} for a summary of limits). 
Our derived limits improve on existing laboratory bounds from other experiments by many orders of magnitude for $m_a \gtrsim 10^{-2}~\textrm{eV}$ (see Fig.~\ref{fig:Results_summary}). 
We note that there are more stringent indirect bounds from the combination of stellar energy-loss arguments and laboratory searches for spin-independent fifth-forces for $m_a \lesssim 10~\textrm{eV}$ \cite{Raffelt2012}, though these bounds may be evaded by certain chameleonic mechanisms, whereby the processes of stellar ``cooling'' due to axion emission become inhibited \cite{Jain2006}. 

Our derived limits also directly constrain \emph{CP} violation in certain types of relaxion models \cite{Graham2015relaxion,Perez2015relaxion,Perez2016relaxion}, where a spin-0 relaxion field $\phi$ couples to the Higgs doublet $H$ via the super-renormalisable interaction $\mathcal{L}_{\phi h h} = -g \phi H^\dagger H$, which induces scalar interactions of $\phi$ with the electron and nucleons \cite{Pospelov2010Higgs}:~$g_e^s = g m_e / m_h^2$ and $g_N^s = g b m_N / m_h^2$, where $m_h$ is the Higgs mass, and the parameter $b \sim 0.2 - 0.5$ \cite{Shifman1978b}. 
Our results constrain the combination of parameters $g g_e^p$ via the relation:
\begin{equation}
\label{relaxion_constraints_CP}
\left| g g_e^p \right|_\textrm{limit} =  \left( \frac{m_h^2}{m_e + b m_N} \right) \left| g_e^s g_e^p + g_N^s g_e^p \right|_\textrm{limit}  \, .
\end{equation}

Finally, we mention that ongoing and future EDM experiments with atoms and molecules (see, e.g., Ref.~\cite{Roberts2015ReviewPNC} for an overview) may improve on the level of sensitivity demonstrated in the present work.

\textbf{Acknowledgements.} --- 
We are grateful to Maxim Pospelov for helpful discussions. 
This work was supported in part by the Australian Research Council. 
Y.~V.~S.~was supported by the Humboldt Research Fellowship. 
V.~V.~F.~was supported by the Gutenberg Research College Fellowship. 
We thank the Mainz Institute for Theoretical Physics (MITP) for its hospitality and support.

\textbf{Note added after publication.} --- 
We have corrected for an underestimation of our published limits \cite{Stadnik2018PRL} on $g_N^s g_e^p X_r/m_a^2$, $g_N^s g_e^p$, $g_e^s g_e^p / m_a^2$ and $g_e^s g_e^p$ from molecular EDM experiments by the factor of $A/Z$. 
The corrected limits are presented in Table~\ref{tab:table2} and Fig.~\ref{fig:Results_summary} [note that our published limits on $g_e^s g_e^p \ln(m_a/m_e) / m_a^2$ remain unchanged]. 
We had unwittingly assumed the commonly-used normalisation factor of $A$ for the contact scalar-pseudoscalar interaction, whereas the relevant molecular calculations use the different normalisation factor of $Z$. 
In Table~\ref{tab:table2} and Fig.~\ref{fig:Results_summary}, we have also added new and improved limits based on data from the newer ThO experiment in \cite{ACME2018ThO}, which appeared after the publication of our present paper \cite{Stadnik2018PRL}.


\section*{Supplemental material}



\textbf{Interactions and potentials.} --- 
One may write the couplings of an axion $a$ with the SM fermions $\psi$ in the following form:
\begin{equation}
\label{general_formula_SP_SUPP}
\mathcal{L}_\textrm{int} = a \sum_\psi \bar{\psi} \left(g_\psi^s + ig_\psi^p \gamma_5 \right) \psi \, .
\end{equation}
The \emph{P},\emph{T}-violating potential due to the exchange of an axion of mass $m_a$ between two fermions reads: 
\begin{equation}
\label{relativistic_potential_FULL_SUPP}
V_{12}(r) = +i \frac{g_1^p g_2^s}{4 \pi} \frac{e^{-m_a r}}{r} \gamma^0 \gamma_5 \, ,
\end{equation}
where $r$ is the distance between the two fermions, and the $\gamma$-matrices correspond to fermion 1. 
The non-derivative form of the potential (\ref{relativistic_potential_FULL_SUPP}) is convenient for performing numerical calculations (see the main text for more details). 
For analytical estimates, it is convenient to use the equivalent derivative form (obtained by using the relation $i m_\psi a \bar{\psi} \gamma_5 \psi = - (\partial_\mu a) \bar{\psi} \gamma^\mu \gamma_5 \psi / 2$) of the potential: 
\begin{equation}
\label{relativistic_potential_SUPP}
V_{12}(r) \approx \frac{g_1^p g_2^s}{8 \pi m_1} \v{\Sigma} \cdot \v{\hat{r}} \left(\frac{m_a}{r} + \frac{1}{r^2} \right) e^{-m_a r} \, ,
\end{equation}
where $m_1$ is the mass of fermion 1, $\v{\Sigma} = \bigl( \begin{smallmatrix}\v{\sigma} & 0\\ 0 & \v{\sigma}\end{smallmatrix}\bigr)$ is the Dirac spin matrix vector of fermion 1, and $\v{\hat{r}}$ is the unit vector directed from fermion 2 to fermion 1.
We restrict our attention to the case when fermion 1 is the electron, but fermion 2 can be either the electron or nucleons. 
We also introduce the shorthand notation $g_N^s \equiv (N g_n^s + Z g_p^s) / A$, where $N$ is the neutron number, $Z$ is the proton number, and $A = Z + N$ is the nucleon number. 
The \emph{P},\emph{T}-violating potentials in Eqs.~(\ref{relativistic_potential_FULL_SUPP}) and (\ref{relativistic_potential_SUPP}) induce EDMs in atoms and molecules by mixing atomic states of opposite parity.


\textbf{Exchange of high-mass axion-like particle.} --- 
When the Yukawa range parameter $\lambda = 1/m_a$ is small compared with the radius of the $1s$ atomic orbital $r_{1s} = a_{\textrm{B}}/Z$ ($a_{\textrm{B}} = 1/m_e \alpha$ denotes the atomic Bohr radius, where $m_e$ is the electron mass and $\alpha \approx 1/137$ is the electromagnetic fine-structure constant), the interaction becomes contact-like. 
For $Z \sim 80$, the corresponding range of axion masses is $m_a \gg 300~\textrm{keV}$. 

\emph{Electron--nucleon interaction.} --- 
We begin by considering the exchange of axions between atomic electrons and nucleons. 
In the contact limit, the effects are dominated by the atomic wavefunctions near the nucleus, where relativistic effects are important. 
The most important matrix elements to consider are, therefore, between $s_{1/2}$ and $p_{1/2}$ atomic states. 
For $r \ll a_{\textrm{B}}/Z^{1/3}$, the relativistic radial wavefunctions corresponding to the upper and lower components take the following respective forms \cite{Khriplovich1991Book}:  
\begin{align}
f_{njl}(r) &= \frac{\kappa}{|\kappa|} \left( \frac{1}{Za_{\textrm{B}} \nu^3} \right)^{1/2} \frac{\left[ (\gamma + \kappa) J_{2\gamma}(x) - \frac{x}{2}J_{2\gamma - 1}(x) \right] }{r} \, , \label{Bessel_upper} \\
g_{njl}(r) &= \frac{\kappa}{|\kappa|} \left( \frac{1}{Za_{\textrm{B}} \nu^3} \right)^{1/2} \frac{Z \alpha J_{2\gamma}(x)}{r} \, \label{Bessel_lower} ,
\end{align}
where $J$ is the Bessel function of the first kind, $x=\sqrt{8Zr/a_{\textrm{B}}}$, $\gamma = \sqrt{(j+1/2)^2 -(Z \alpha)^2}$, $\kappa = (-1)^{j+1/2-l}(j+1/2)$, and $\nu$ is the effective principal quantum number. 

Using the wavefunctions (\ref{Bessel_upper}) and (\ref{Bessel_lower}), together with either form of the operator (\ref{relativistic_potential_FULL_SUPP}) or (\ref{relativistic_potential_SUPP}) with $m_a \gg Z \alpha m_e$, 
we calculate the relevant matrix element to be: 
\begin{equation}
\label{relativistic_ME_contact}
\lim_{m_a \to \infty} \left< ns_{1/2} \left| V_{eN} \right| n'p_{1/2} \right> = \frac{A g_N^s g_e^p Z^2 \alpha^4 m_e^3 \gamma K_r}{2\pi m_a^2 \left( \nu_{ns} \nu_{n'p}\right)^{3/2}} \, ,
\end{equation}
where the relativistic factor $K_r$ is given by:
\begin{equation}
\label{contact_rel_factor}
K_r = \left[ \frac{2}{\Gamma(2\gamma +1)} \left( \frac{2Z r_c}{a_{\textrm{B}}} \right)^{\gamma-1} \right]^2 \, .
\end{equation}
The cut-off radius $r_c$ is given by $r_c \approx R_{\textrm{nucl}} \approx 1.2 A^{1/3}~\textrm{fm}$ when $m_a R_{\textrm{nucl}} \gg 1$, and $r_c \approx 1/m_a$ when $m_a R_{\textrm{nucl}} \ll 1$, where $R_{\textrm{nucl}}$ is the radius of the atomic nucleus. 

It is convenient to relate the matrix element (\ref{relativistic_ME_contact}) to the corresponding matrix element of the generic contact interaction $\mathcal{L}_{\textrm{int}}^{\textrm{contact}} = - G_F C_\textrm{SP} \bar{N}N \bar{e}i\gamma_5 e / \sqrt{2}$, which reads \cite{Khriplovich1991Book}:
\begin{equation}
\label{generic_ME_contact}
\left< ns_{1/2} \left| V_{\textrm{int}}^{\textrm{contact}} \right| n'p_{1/2} \right> = - \frac{A G_F C_{\textrm{SP}} Z^2 \alpha^4 m_e^3 \gamma K_r'}{2 \sqrt{2} \pi \left( \nu_{ns} \nu_{n'p}\right)^{3/2}} \, ,
\end{equation}
where $G_F$ is the Fermi constant, and $K_r'$ is the relativistic factor (\ref{contact_rel_factor}) with the cut-off radius given by $r_c = R_{\textrm{nucl}}$. 
This allows us to make use of numerical calculations, which relate $C_\textrm{SP}$ to the induced EDM in atoms and to the \emph{P},\emph{T}-odd spin-axis interaction in molecules. 
Comparing (\ref{relativistic_ME_contact}) with (\ref{generic_ME_contact}), we find that:
\begin{equation}
\label{C_SP-effective_eN}
C_{\textrm{SP}}^{\textrm{equiv}} = - \frac{\sqrt{2} g_N^s g_e^p}{G_F m_a^2} X_r \, ,
\end{equation}
where $X_r \approx 1$ when $m_a R_{\textrm{nucl}} \gg 1$, and $X_r \approx (m_a R_{\textrm{nucl}})^{2-2\gamma}$ when $m_a R_{\textrm{nucl}} \ll 1$. 


\emph{Electron--electron interaction.} --- 
When high-mass axions are exchanged between atomic electrons, the valence atomic electrons now interact predominantly with a `core' of two $1s$ electrons (which are situated mainly at the distances $r \sim r_{1s} = a_{\textrm{B}}/Z$), instead of with the $A$ nucleons of the nucleus. 
We estimate the relevant non-relativistic matrix element:
\begin{equation}
\label{NRL_ME_ee_contact}
-i \frac{g_e^s g_e^p}{2m_e m_a^2} \int  n_e(\v{r}) \psi_{ns}^\dagger(\v{r}) \left( \v{\sigma} \cdot \v{\overrightarrow{p}} - \v{\sigma} \cdot \v{\overleftarrow{p}} \right) \psi_{n'p}(\v{r}) ~ d^3r \, , 
\end{equation}
where $n_e$ is the number density of electrons and $\v{p}$ is the electron momentum operator, by using the non-relativistic limit of the radial wavefunction (\ref{Bessel_upper}) for the valence electron and the non-relativistic hydrogen-like Coulomb wavefunction for the $1s$ electrons:
\begin{equation}
\label{relativistic_ME_contact_ee}
\lim_{m_a \to \infty} \left< ns_{1/2} \left| V_{ee} \right| n'p_{1/2} \right> = \frac{K_{1s} g_e^s g_e^p Z^2 \alpha^4 m_e^3}{\pi m_a^2 \left( \nu_{ns} \nu_{n'p}\right)^{3/2}} \, ,
\end{equation}
where $K_{1s} = [2I_1(2) - I_0(2)]/e^2 \approx 0.122$ is a constant, with $I$ being the modified Bessel function of the first kind and $e \approx 2.72$ being Euler's number. 
Comparing (\ref{relativistic_ME_contact_ee}) with (\ref{generic_ME_contact}), we find that:
\begin{equation}
\label{C_SP-effective_ee}
C_{\textrm{SP}}^{\textrm{equiv}} = - \frac{2 \sqrt{2} K_{1s} g_e^s g_e^p}{A G_F m_a^2 \gamma K_r'}  \, .
\end{equation}

The contribution of the interaction of the valence atomic electrons with non-$1s$ electrons is parametrically suppressed, scaling only as $\propto Z^{5/3}$ from a semi-classical treatment of the matrix element (\ref{NRL_ME_ee_contact}). 

\textbf{Exchange of low-mass axion-like particle.} --- 
When the Yukawa range parameter $\lambda = 1/m_a$ is large compared with the radius of the atom $R_{\textrm{atom}}$, the interaction becomes long-range. 
For heavy atomic species, which are of experimental interest, $R_{\textrm{atom}} \approx 4 a_{\textrm{B}}$, and so the corresponding range of axion masses is $m_a \ll 1~\textrm{keV}$. 

\emph{Electron--nucleon interaction.} --- 
We again begin by considering the exchange of axions between atomic electrons and nucleons. 
In the limit as $m_a \to 0$, the operator (\ref{relativistic_potential_SUPP}) takes the form (after summation over the nucleons):
\begin{equation}
\label{relativistic_potential_massless}
\lim_{m_a \to 0} V_{eN}(r) = \frac{Ag_N^s g_e^p}{8 \pi m_e} \frac{\v{\Sigma} \cdot \v{\hat{r}}}{r^2} \, .
\end{equation}
We can estimate the matrix elements of the operator (\ref{relativistic_potential_massless})
semi-classically: 
\begin{equation}
\label{ME_eN_low-mass_TF}
\lim_{m_a \to 0} \left| \left< n, l = j-1/2 \left| V_{eN} \right|  n', l = j+1/2 \right> \right| \sim \frac{A |g_N^s g_e^p|}{8 \pi m_e a_\textrm{B}^2} \, . 
\end{equation}

Also, with the aid of the identity $\v{\Sigma} \cdot \v{E} = [\v{\Sigma} \cdot \v{\nabla}, H_{\textrm{Dirac}}]/e$, where $\v{E} = \v{E}_\textrm{int} + \v{E}_\textrm{ext}$ is the sum of the internal and external electric fields, and  $H_{\textrm{Dirac}}$ is the relativistic Dirac atomic Hamiltonian, we can write the ``residual'' non-vanishing (after summation over all intermediate opposite-parity atomic states) part of the operator (\ref{relativistic_potential_massless}) as follows (keeping only terms that produce a linear atomic energy shift in an external electric field):
\begin{equation}
\label{relativistic_potential_massless_residual}
\lim_{m_a \to 0} V_{eN}^\textrm{residual}(r) = \frac{Ag_N^s g_e^p}{8 \pi m_e} \left[ \frac{\v{\Sigma} \cdot \v{\hat{r}}}{r^2} - \frac{\v{\Sigma} \cdot \v{E}_\textrm{int}}{e} \right] \, .
\end{equation}

Using the relativistic radial wavefunctions (\ref{Bessel_upper}) and (\ref{Bessel_lower}), it is straightforward to verify that the contribution to the matrix element $\lim_{m_a \to 0} \left< n, l = j-1/2 \left| V_{eN}^\textrm{residual} \right|  n', l = j+1/2 \right>$ from the small distances, $r \ll a_{\textrm{B}}/Z^{1/3}$, vanishes. 
Similarly, the contribution to this matrix element from the large distances, $r \gg a_{\textrm{B}}/Z^{1/3}$, also vanishes, since $\v{E}_\textrm{int} = e\v{\hat{r}}/r^2$ at large distances. 
The non-vanishing contribution to this matrix element, therefore, arises at the intermediate distances, $r \sim a_{\textrm{B}}/Z^{1/3}$. 
This is in contrast to the case of the contact interaction, where the dominant contribution to the relevant matrix elements comes from the small distances, $r \ll a_{\textrm{B}}/Z^{1/3}$. 

Since the effects are not dominated by the atomic wavefunctions near the nucleus in this case, the contributions from higher angular-momentum atomic states are not necessarily suppressed (in contrast to the case of the contact interaction, where the $j=1/2$ atomic states dominate).

\emph{Electron--electron interaction.} --- 
In the case of the exchange of low-mass axions between atomic electrons, the main contribution arises from the interaction of the valence atomic electrons with non-$1s$ electrons. 
We again treat the relevant non-relativistic matrix element,
\begin{equation}
\label{ME_ee_low-mass}
\frac{g_e^s g_e^p}{8 \pi m_e} \iint  n_e(\v{r}_2) \psi_{A}^\dagger(\v{r}_1) \frac{\v{\sigma}_1 \cdot \v{\hat{r}}_{12}}{r^2_{12}} \psi_{B}(\v{r}_1) ~ d^3r_1 d^3r_2 \, , 
\end{equation}
semi-classically. 
From the comparison of the integrals in (\ref{ME_eN_low-mass_TF}) and (\ref{ME_ee_low-mass}), we see that the two matrix elements are related to each other via the relation:
\begin{equation}
\label{ME_ee_eN_semi-classical_relation}
|g_N^s g_e^p|A \approx |g_e^s g_e^p|Z \, . 
\end{equation}
We hence arrive at the following estimate: 
\begin{equation}
\label{ME_ee_low-mass_TF}
\lim_{m_a \to 0} \left| \left< n, l = j-1/2 \left| V_{ee} \right|  n', l = j+1/2 \right> \right| \sim \frac{Z |g_e^s g_e^p|}{8 \pi m_e a_\textrm{B}^2} \, . 
\end{equation}

\end{document}